\documentclass[12pt]{article}
\usepackage{amsmath}
\usepackage{graphicx,psfrag,epsf}
\usepackage{enumerate}
\usepackage{xcolor}
\usepackage{amsfonts}
\usepackage{algorithm}
\usepackage{algpseudocode}
\usepackage{biblatex}
\usepackage{mathrsfs}
\newtheorem{theorem}{Theorem}

\newtheorem{defi}{Definition}

\newcommand{\blind}{0}

\date{}
\begin{document}

\def\spacingset#1{\renewcommand{\baselinestretch}%
{#1}\small\normalsize} \spacingset{1}


\if0\blind
{
  \title{\bf Community Detection through Recursive Partitioning in Bayesian Framework}
  \author{Yuhua Zhang\thanks{
    yuhuazhang@hsph.harvard.edu}\hspace{.2cm}\\
    Department of Biostatistics, Harvard University\\
     \\
    Kori S. Zachrison \\
    Department of Emergency Medicine, Massachusetts General Hospital\\
     \\
    Renee Y. Hsia \\
    Emergency Medicine and Institute of Health Policy Studies,\\ University of California, San Francisco\\
     \\
    Jukka-Pekka Onnela\hspace{.2cm} \\
    Department of Biostatistics, Harvard University}
  \maketitle
} \fi

\if1\blind
{
  \bigskip
  \bigskip
  \bigskip
  \begin{center}
    {\LARGE\bf Title}
\end{center}
  \medskip
} \fi

\bigskip
\begin{abstract}
Community detection involves grouping the nodes in the network and is one of the most-studied tasks in network science. Conventional methods usually require the specification of the number of communities $K$ in the network. This number is determined heuristically or by certain model selection criteria. In practice, different model selection criteria yield different values of $K$, leading to different results. We propose a community detection method based on recursive partitioning within the Bayesian framework. The method is compatible with a wide range of existing model-based community detection frameworks. In particular, our method does not require pre-specification of the number of communities and can capture the hierarchical structure of the network. We establish the theoretical guarantee of consistency under the stochastic block model and demonstrate the effectiveness of our method through simulations using different models that cover a broad range of scenarios. We apply our method to the California Department of Healthcare Access and Information (HCAI) data, including all Emergency Department (ED) and hospital discharges from 342 hospitals to identify regional hospital clusters. 
\end{abstract}

\noindent%

\spacingset{1.45}
\section{Introduction}
\label{sec:intro}



Network data have become ubiquitous in many fields. Among various network analysis tasks, community detection is one of the most-studied. Community detection consists of clustering network nodes into groups such that nodes in each community are more densely connected internally than externally. In many applications, communities provide a useful representation of the network. In the stroke transfer network, for example, hospital clusters can be identified through community detection in the patient tranfer network. These detected communities reveal regional connection patterns among hospitals and can be further utilized to investigate disparities in health care access.

A popular community detection method is the class of stochastic block models (SBMs) \cite{holland1983stochastic}. The simplest version of the SBM assumes that vertices within the same block have the same probability of forming an edge. Internal connections are more likely than external connections in each group. Multiple extensions of the SBM have been proposed, such as the degree-corrected stochastic block model (DC-SBM)~\cite{karrer2011stochastic, sengupta2018block, noroozi2021estimation, koo2022popularity}, mixed membership stochastic block models~\cite{airoldi2008mixed}, as well as others~\cite{aicher2015learning,  galhotra2018geometric, sengupta2018block, noroozi2021estimation, koo2022popularity, peixoto2018nonparametric, ng2021weighted}. Many inferential methods have been proposed for block models. One large class of these is the spectral clustering methods~\cite{rohe2011spectral, chin2015stochastic}. Another large class deals with model-based likelihood or pseudo-likelihood~\cite{van2018bayesian, morup2012bayesian,amini2013pseudo, strauss1990pseudolikelihood}. In addition to block models, latent space models have been used to discover community structure, where the probability of a tie between two nodes depends on the distance between them in an unobserved latent Euclidean space~\cite{gao2022community, handcock2007model, hoff2002latent}.

Most existing methods require predetermining the number of communities $K$ in the network. Once the value of $K$ is fixed, a single $K$-partitioned block model is fitted to the data. Determination of $K$ is a model selection problem. For model-based methods, popular model selection criteria include information criteria, eigenvalues from spectral clustering, and other likelihood-based criteria \cite{rohe2011spectral, chin2015stochastic,yan2016bayesian,spiegelhalter2014deviance}.  However, in many real-world cases, different model selection criteria lead to different values of $K$, eventually yielding different community structure. In addition, hierarchical structure is often observed in real networks and can be meaningful. For example, in the stroke patient transfer network, communities, which are in part determined by geography and in part by hospital affiliations, are fairly autonomous units in the health care delivery system. To avoid the inconsistency caused by adopting different model selection criteria in determining the value of $K$, as well as to unveil the hierarchical structures of the network, a community detection method that splits the network in a top-down fashion, e.g., a tree-structured recursive partitioning method, is desirable.

Similar ideas have been explored in previous works. One lineage of these methods is the direct modeling of the hierarchical structure in the Bayesian framework~\cite{blundell2013bayesian,clauset2008hierarchical}. These models attempt to infer the tree structure directly but they tend to be computationally intractable for large networks. Furthermore, these methods are typically built upon a specific model and are not generalizable to others. Another lineage of methods uses recursive bipartitioning algorithm based on spectral clustering~\cite{dasgupta2006spectral,balakrishnan2011noise,li2022hierarchical}. In practice, networks display features that are well captured by some models, and spectral clustering methods can fall short of capturing these features.


In this paper, we consider a framework for community detection based on recursive bipartitioning. The two components of our framework are the bipartitioning algorithm and the stopping rule. The algorithm recursively divides the given network into two until it hits the stopping rule. Our proposed framework uses the Bayes factor as the stopping rule: when the data favor a model with community structure over a model without community structure, the Bayes factor is relatively large and the algorithm divides the current network into two. In principle, any model-based community detection method with a likelihood can be used in our framework. Our main contribution is a new community detection method in the Bayesian framework that is computationally tractable and does not require pre-specification of the number of communities. In addition, our proposed framework allows modeling of network properties through likelihood functions. 

The rest of the paper is organized as follows. In Section 2, we introduce our notation and models. We also discuss the consistency of the recursive bipartitioning framework under the stochastic block model in this section. In Section 3, we discuss the inferential algorithm. In Section 4, we demonstrate our method in a simulation study. In Section 5, we apply our proposed method to a real-world stroke patient transfer data set. We conclude with a discussion in Section 6. 


\section{Model Description}
\label{s:notation}

\subsection{Background}
Network data consists of nodes and edges. Nodes represent entities in the network and edges represent interactions among these entities. The community detection problem can be formulated as finding a disjoint partition of nodes, where each of these node set is called a community.
Let~$K$ be the number of communities. The true value of $K$ is unknown in the observed networks. Under different models, the community detection are formulated in different ways. We focus on three classes of network models in this paper: stochastic block models, edge exchangeable models, and latent space models, based on which we will demonstrate our proposed method. In this section, we will go through notations and assumptions in these three models. 

\noindent\textbf{Stochastic Block Model:}\quad The network is denoted as $Y=(V,E)$, where $V$ is the set of nodes, $E$ is the set of edges, and $n=|V|$ is the number of nodes. The network can be represented by an $n\times n$ adjacency matrix $A=[A_{ij}]$, where $A_{ij}=1$ if there is an interaction between nodes $i$ and $j$, and otherwise $A_{ij}=0$. In the SBM, the network is assumed to be undirected and nodes in the same community are assumed to have the same connection probability such that the expended value of $A_{ij}$ is given by
$$\mathbf{E}(A_{ij})=\mathcal{B}(B(i),B(j)),$$
where $\mathcal{B}$ is a $K$ by $K$ matrix  and $B(\cdot)$ is the community assignment of each node. Given the $\mathcal{B}$ matrix, it is commonly assumed that $A_{ij}\sim \text{Bernoulli}(\mathcal{B}(B(i),B(j)))$ independently for each pair of $i$ and $j$. This formulation allows for self-loops.

\noindent\textbf{Edge Exchangeable Model:}\quad In the edge exchangeable model, the network $Y$ is assumed to be random. Let $\mathcal{P}$ denote the countable node population. The network is defined as an interaction process $\mathcal{I}:\mathbb{N}\rightarrow \mathcal{P}$. Let $E_j=(S_j, R_j)$ denote a single directed interaction in the network with a sender $S_j$ and a receiver $R_j$, where both $S_i$ and $R_i$ are random samples from $\mathcal{P}$. Let $\{f_i : i\in\mathcal{P}, \sum_{i\in\mathcal{P}}f_i=1\}$ denote the limiting frequencies of classes, which can be interpreted as the set of propensities to observe nodes in the network. 

Community detection in the edge exchangeable model is to find a partition of the nodes such that $\mathcal{P}=\{\mathcal{P}_1,...,\mathcal{P}_K\}$. Denote the node population in block $b$ as $\mathcal{P}_b$. The nodes in different blocks are assumed to be disjoint, i.e., $\mathcal{P}_b\cap\mathcal{P}_{b'}=\emptyset,\forall b\neq b'$. 
Similar to SBMs, it is assumed that nodes in the same community have the same probability of connecting to other nodes in the network.
In a directed network, the probability of a pairwise interaction $E_m$ is given by
$$P(E_j=\{S_j,R_j\})=\pi_{B(S_j)}f_{S_j}\mathcal{B}(B(S_j),B(R_j))f_{R_j},$$
where $\mathcal{B}$ is a $K$ by $K$ matrix with each entry corresponding to the probability of an interaction initiated from one block (row) to another block (column) with $\sum_{k=1}^{K}\mathcal{B}(k',k)=1$, and $\pi$ is a $K$-vector indicating the probability of an interaction being initiated from each block. 



\noindent\textbf{Latent Space Model:} In the latent space model, the network is assumed to be binary and represented by an $n$ by $n$ matrix $\mathbf{Y}$, where entry $Y_{ij}$ indicates whether the nodes $i$ and $j$ are connected. Suppose each node has an unobserved position in a $d$-dimensional latent Euclidean space denoted by $Z=\{z_i\}$. The probability that $Y_{ij}=1$ is given by the generalized linear model
$$\text{logit}P(Y_{ij}=1|z_i,z_j,x_{ij},\beta)=\beta_0x_{ij}-\beta_{1}|z_i-z_j|,$$
where the $\beta$s are regression coefficients, $|z_i-z_j|$ is the Euclidean distance between nodes $i$ and $j$ in the latent space and $x_{ij}$ denotes observed attributes that may be dyad specific. 
To incorporate community structure in the latent space model, one approach is to assume $z_i$ is drawn from a mixture of  $d$-dimensional multivariate normal distributions, such that
$$z_i\sim \sum_{k=1}^{K}\rho_k\text{MVN}_d(\mu_k,\sigma_k^2I_d),$$
where $\rho_k$ is the probability of a node belonging to community $k$ such that $\sum_{k=1}^{K}\rho_k=1$, and $\text{MVN}_d(\mu_k,\sigma_k^2I_d)$ is the $d$-dimensional multivariate Gaussian distrubution with $\mu_k$ and $\sigma_k^2$ being the mean and variance of the $k^{\text{th}}$ component of the Gaussian mixture. 

\subsection{Recursive Bipartitioning Method}


We adopt a recursive partitioning framework in this paper. We refer to the process of dividing the current network into two smaller communities as bipartitioning. Recursive bipartitioning has two components: (1) an algorithm that can effectively bipartition the network, and (2) a stopping rule to determine whether community structure exists in the current network. The algorithm stops dividing the network if the condition in the stopping rule is met. 
Intuitively, the algorithm starts by dividing the network into large, loosely connected communities and then proceeds to break them down into small, densely connected communities until no more than one community remains. 

Denote the complete network as $Y=(V,E)$, where $V$ is the set of nodes and $E$ is the set of edges. The bipartitioning yields two sub-networks $Y_1=(V_1,E_1)$ and $Y_2=(V_2,E_2)$, where $V_1\cap V_2=\emptyset$ and $V_1\cup V_2=V$. Denote $E_{-12}=E\backslash (E_1\cup E_2)$, which represents the interconnections between the two sub-networks $Y_1$ and $Y_2$. Furthermore, $Y_1$ and $Y_2$ will be used as input for the subsequent rounds of bipartitioning.



The process stops when the remaining sub-networks are no longer separable, resulting in a disjoint set of nodes $\{V_1,...,V_K\}$, where $V_1 \cup ...\cup V_K=V^{(0)}$. A key criterion of any community detection framework is whether the method can recover the true underlying labels, that is, the consistency of the inferred community labels with the true labels. In our case, it refers to whether the disjoint set of nodes given by the bipartitioning algorithm represents the true communities. 
With a slight abuse of notation, denote the disjoint set of nodes $\{V_1,...,V_K\}$ as the partition of the nodes into the true communities, such that $\forall i\in V_k, B(i)=k$. 
Numerous consistency results have been proven for a variety of community detection models assuming the number of communities is known. We show that results given by the recursive bipartitioning method are consistent in the stochastic block model when the number of communities is unknown.  
The consistency often comes with specific criteria and models. One widely accepted criterion in the stochastic block models is the Erd\H{o}s-Renyi modularity (ERM),
$$Q_{ERM}(e)=\sum_{k=1}^{K}(O_{kk}-\frac{n^{2}_{k}}{n^2}L),$$
where $e$ is an arbitrary set of community labels, $O_{kk}=\sum_{ij}A_{ij}I\{e_i=k,e_j=k\}$, $n_k$ is the number of nodes in each community under $e$, and $L=\sum_{ij}A_{ij}$. 
Given the criterion, we formally define strong and weak consistency of a community detection method.


\begin{defi}(Consistency)
A community detection criterion $Q$ is said to be strongly consistent if the node labels obtained by maximizing the criterion, i.e., $\{\hat{B}(i),i\in [n]\}=\arg\max_e Q(e)$, satisfy
$$P(\hat{B}(i)=B(i))\rightarrow 1\text{\quad as\ } n\rightarrow\infty.$$

A community detection criterion $Q$ is said to be weakly consistent if the node labels obtained by maximizing the criterion satisfy
$$\forall \epsilon>0, P\left[ \left(\frac{1}{n}\sum_{i=1}^{n}1(\hat{B}(i)\neq B(i))\right)<\epsilon\right]\rightarrow 1\text{\quad as\ } n\rightarrow\infty.$$

\end{defi}



One assumption that comes with the ERM is that $P_{aa}>P_0$ and $P_{ab}<P_0$, where $a,b\in [K]$, $P_0=\sum_{ab}\pi_a\pi_bP_{ab}$, and $\pi_a=P(e_i=a)$. In other words, the probability of a connection within the same community is higher than the probability of the connection between communities. 

Denote the subnetwork as $Y_S=(V_S,E_S)$, where $V_S=\cup_{j\ge 1,j\in [K]}V_j$ and its biparitioned subnetworks as $\{Y_{S_1},Y_{S_2}\} = \{(V_{S_1},E_{S_1}), (V_{S_2},E_{S_2})\}$, where $V_{S_1}=\cup_{k\ge 1,k\in [K]}V_k$ and $V_{S_2}=\cup_{l\ge 1,l\in[K]}V_l$ are two disjoint subsets of $V_S$. Denote $\tilde{P}_{11}=E(A_{ij}),i,j\in V_{S_1}$, $\tilde{P}_{12}=E(A_{ij}),i\in V_{S_1}, j\in V_{S_2}$, $\tilde{P}_{21}=E(A_{ij}),i\in V_{S_2},j\in V_{S_1}$, and $\tilde{P}_{22}=E(A_{ij}),i,j\in V_{S_2}$. Assume $\tilde{P}$ is symmetric, then $\tilde{P}_{12}=\tilde{P}_{21}$. We make a similar assumption to that in the ERM: $P_{aa}>P_0$ and $P_{ab}<P_0$ (for $a\neq b$), where $a,b\in[K]$. This leads to the following conclusion.

\begin{theorem}
\label{thm:PP}
    Suppose that $P_{aa}>P_0>P_{ab}$, for all $a\neq b$. Then for any sub-network $Y_S=\{V_{S},E_S\}$, where $V_S=\cup_{k\ge 2,k\in [K]}V_k$, there exists at least one bipartition of $Y_S$ into $\{Y_{S_1},Y_{S_2}\}$ such that $\tilde{P}_{aa}>\tilde{P}_{ab}$ for all $a\neq b$.
\end{theorem}



Theorem~\ref{thm:PP} shows the existence of at least one partition of an arbitrary sub-network that satisfies the assumptions required for ERM consistency. This leads to consistency of the recursive bipartitioning method under ERM. 

\begin{theorem}
\label{th:cons}

In the stochastic block model, the model criterion $Q_{ERM}$ is strongly consistent when $\lambda_{n_s}/\log {n_s}\rightarrow \infty$ and is weakly consistent when $\lambda_{n_s}\rightarrow\infty$, where $\lambda_{n_s}$ is the expected degree of the sub-network $Y_S$ and $n_s$ is the number of nodes in $Y_S$.

\end{theorem}

Theorem~\ref{th:cons} establishes the theoretical foundations for the consistency of bi-partitioning on any sub-network $Y_S$.
Given that the bipartitioning divides the network recursively until no more than one community exists,
it suffices to show the consistency of the recursive bipartitioning in the stochastic block model under ERM. 
Though we do not provide theoretical consistency guarantees for the edge exchangeable model or the latent space model, our results demonstrate the performance of the proposed framework when applied to the simulation data in Section~\ref{s:simu}.

\subsection{Partitioning Algorithm and Stopping Rule}

We adopt the Bayesian framework for the partitioning algorithm and the stopping rule. 
The task of the partitioning algorithm initially is to identify the community structure when $K=2$. In the stochastic block model and the edge exchangeable model, this means that the $\mathcal{B}$ matrix is a 2 by 2 matrix, with the diagonal entries being the probabilities of within community connections, and the off-diagonal entries being the probability of an inter-community connection in the network. In the latent class model, there are two components in the Gaussian mixture distribution. We utilize MCMC as the partitioning algorithm in this paper.


We use the Bayes factor as the stopping rule to measure the strength of evidence for one model over another by comparing the likelihoods of the two models.  Similar to the idea of the frequentist hypothesis testing, we compare the likelihood of the models with ($M_1$) and without ($M_0$) community structure. If the data support the model with community structure, the algorithm will divide the network. Let

$$BF_{01}=\frac{P(Y|M_1)}{P(Y|M_0)},$$
and let $\mathcal{A}_0$ and $\mathcal{A}_1$ be the parameter spaces that are directly related to community structure under $M_0$ and $M_1$, respectively. One such example of is the $\mathcal{B}$ matrix in block models; another example is the means $\{\mu_k\}$ of the Gaussian mixtures in the latent space model. We assume $\mathcal{A}_0\cup\mathcal{A}_1=\Psi$ and $\mathcal{A}_0\cap\mathcal{A}_1=\emptyset$, where $\Psi$ is the entire parameter space. Let $a_0\in\mathcal{A}_0$ and $a_1\in\mathcal{A}_1$. Let $\Theta_{i,-a_i}$ denote the other parameters in the model $M_i$. The marginal likelihood can be expanded as 
$$P(\mathbf{Y}|M_i)=\int\int P(\mathbf{Y}|a_i,\Theta_{i,-a_i}, M_i) P(\Theta_{i,-a_i}|M_0)P(a_i|M_i)\ \,d\Theta_{i,-a_i}\,da_i.$$


\section{Inference}
\label{s:inf}

The two components in our tree-structured partitioning  algorithm are the recursive division and the stopping rule. We derive a Gibbs-sampling algorithm for the three models and demonstrate how the Bayes factor can be used as the stopping rule in this section. 

\subsection{Gibbs Sampling Algorithm} 

The pseudo-code of the Gibbs sampling algorithms for the three model classes are shown in Algorithm~\ref{alg:sbm}, Algorithm~\ref{alg:cap}, and Algorithm~\ref{alg:lsm}. In the stochastic block model, 
the likelihood of the network $Y$ is given by the product over all pairs of nodes given $\{B(i),i\in [n]\}$ and the $\mathcal{B}$ matrix:
\begin{equation}
\label{eq:1}
\begin{split}
P(Y|\{B(i),i\in [n]\},\mathcal{B})&=\prod_{i\neq j}P(A_{ij}|\{B(i),i\in [n]\},\mathcal{B})\\
&=\prod_{i\neq j}\mathcal{B}(B(i),B(j))^{A_{ij}}(1-\mathcal{B}(B(i),B(j)))^{1-A_{ij}}.
\end{split}
\end{equation}

Note that under this formulation, we exclude self-loops. We assume the following priors:
$$\mathcal{B}(k,k')\sim U(0,1),\ k=1,...,K;\ k'=1,...,K$$
$$P(B(i)=k|\pi)=\pi_k, k=1,...,K;\ \pi\sim \text{Dirichlet}(\gamma_1,...,\gamma_K).$$

Based on these prior specifications, the Gibbs sampling algorithm for sampling the posterior distributions is given in Algorithm~\ref{alg:sbm}.

\begin{algorithm}
\caption{Gibbs sampling algorithm for the stochastic block model}
\label{alg:sbm}
\begin{algorithmic}
\State Denote: 
\begin{align}
n_k&=\sum_{i=1}^{n}I(B(i)=k),\ k=1,...,K\ \nonumber\\
n_{kk'}&=\sum_{i\neq j}I(B(i)=k,B(j)=k')=n_kn_{k'}-n_kI(k=k')\nonumber \\
A[kk']&=\sum_{i,j:B(i)=k,B(j)=k'}A_{ij},\ k=1,...,K,\ k'=1,...,K\nonumber 
\end{align}
\For{\text{the iteration $l$}}
    \State Update $\pi$ and $\mathcal{B}$ according to:
    \begin{align}
    \pi|\gamma, n_{[k],k=1,...,K}&\sim \text{Dirichlet}(\gamma_1+n_1,...,\gamma_K+n_K)\nonumber \\
    \mathcal{B}|n_{[kk'],k,k'=1,...,K},A&\sim \text{Beta}(1+A[kk'],1+n_{kk'}-A[kk'])\nonumber 
    \end{align}
    \State Update the terms involves $B(i)$ for each node:
    $$P(B(i)=k|\{B(j),j\neq i\},A,\pi,\mathcal{B})\propto\pi_k\times \{\prod_{j\neq i}\mathcal{B}(k,B(j))^{A_{ij}}(1-\mathcal{B}(k,B(j)))^{1-A_{ij}}\}$$
    $$\times \{\prod_{l\neq i}\mathcal{B}(B(l),k)^{A_{li}}(1-\mathcal{B}(B(l),k))^{1-A_{li}}\}$$
\EndFor
\end{algorithmic}
\end{algorithm}

\begin{algorithm}
\caption{Gibbs sampling algorithm for the edge exchangeable model}\label{alg:cap}
\begin{algorithmic}
\State Denote:
    \begin{align}
    m^k(Y_M)&=\sum_{i,B(i)=k}k_i(M),\text{\ where\ }k_i(M) \text{\ is\ the degree of node\ } i \nonumber \\
    D_M^{k}(i)&=\sum_{m=1,S_m=i}^{M}I(B(R_m)=k)\nonumber 
    \end{align}
\For{\text{the iteration $l$}}
        \State Update $B(i)$ for each node $i\in[n]$ according to Eq (\ref{eq:c})
    \For{$k\in [K]$}
        \State Update $\alpha_k$ using auxiliary variables according to Eq (\ref{eq:alpha_c})
        \State Update $\theta_k$ using auxiliary variable according to Eq (\ref{eq:theta_c})
    \EndFor
\EndFor
\end{algorithmic}
\end{algorithm}

In the edge exchangeable model, suppose the network is composed of directed interactions involving two nodes, the likelihood of the model is:
\begin{equation}
\label{eq:2}
P(Y_M|\mathcal{B},\{B(i),i\in\mathcal{P}\},\pi,\{f_i,i\in\mathcal{P}\})=\prod_{m=1}^{M}\pi_{B(S_m)}f_{S_{m}}\times\mathcal{B}(B(S_m),B(R_m)) f_{R_m}
.
\end{equation}

In order to derive the Gibbs sampling algorithm, we need to specify a representation of $f$, such that $f_i = W_i \prod_{i' < i} (1 - W_{i'})$, where $W_i$ are independent and $W_i\sim\text{Beta}(1-\alpha,\theta+i\alpha)$. This is also known as the Griffiths–Engen–McCloskey (GEM) distribution with parameters $(\alpha, \theta)$. To incorporate community structure, we assume a block-specific representation of $f$, where $f_i^{(b)} = W_i \prod_{i' < i} (1 - W_{i'})$ and $W_i\sim\text{Beta}(1-\alpha_b,\theta_b+i\alpha_b)$. $(\alpha_b,\theta_b)$ denote block-specific model parameters.

We first update the block assignments $\{B (i),i\in\mathcal{P}\}$.  The probability $p_k$ of assigning node $i$ to block $k\in [K]$, given the other model parameters, is
$$
p_k := P(B(i)=k)\times P(Y_M|\mathcal{B},\{B(j),j\neq i,j\in\mathcal{P}\},\pi,\{f_i,i\in\mathcal{P}\}).
$$
Similar to the SBM, we assume $P(B(i)=k|\pi)=\pi_k$ and $\pi\sim \text{Dirichlet}(\gamma_1,...,\gamma_K)$. The probability of assigning node $i$ to community $k$ follows a multinomial distribution with the probability of category-$k$ being
\begin{equation}
\label{eq:c}
P(B(i)=k|Y_M)=\frac{p_k}{\sum_{k'=1}^{K}p_{k'}}.
\end{equation}

The conditional updates of $\{\alpha_k\}$ and $\{\theta_k\}$ proceed as follows. The values of $\alpha_k$ and $\theta_k$ are sampled from the following posterior distributions using the auxiliary variable method proposed by \cite{teh2006bayesian}:
\begin{align}
x_k &\sim \text{Beta} \left(\theta_{k}+1,m^{k}(\mathbf{Y}_{M})-1 \right), k \in [K] \nonumber \\
y_{i,k}&\sim \text{Bernoulli} \left( \frac{\theta_{k}}{\theta_{k}+\alpha_k \cdot i} \right),  
i =1,\ldots,n_k-1, k\in[K] \nonumber \\
z_{i,j,k} &\sim \text{Bernoulli} \left( \frac{j-1}{j-\alpha_k} \right),  
i=1,\ldots,n_k, j = 1,\ldots, D_M^{k} (i)-1 \nonumber \\
\theta_k &\sim \text{Gamma} \left(\sum_{i=1}^{n_k-1}y_{i,k}+a,b-\log x_k \right), \label{eq:theta_c} \\
\alpha_k &\sim \text{Beta}\left(c+\sum_{i=1}^{n_k-1}(1-y_{i,k}),d+\sum_{i=1}^{n_k} \sum_{j=1}^{D_M^{k}(i)-1}(1-z_{i,j,k}) \right), \label{eq:alpha_c}
\end{align}
where~$\{a,b,c,d\}$ are the parameters of the Gamma and Beta priors, and $\{x,y,z\}$ are auxiliary variables. We assume $\theta_k\sim \text{Gamma}(a,b)$ and $\alpha_k\sim \text{Beta}(c,d)$ as priors.

Lastly, in Algorithm~\ref{alg:lsm} for the latent space model, the likelihood of the model is given by:

\begin{equation}
\label{eq:3}
P(Y|\beta,Z,X)=\prod_{i\neq j}P(Y_{ij}|z_i,z_j,x_{ij},\beta).
\end{equation} 

We specify the following prior distributions:
\begin{align}
\beta&\sim \text{MVN}_p(\xi,\Psi),\ \lambda\sim\text{Dirichlet}(\nu)\nonumber \\
\sigma_k^2&\sim\sigma_0\text{Inv}\chi_{\alpha}^2,\ k=1,...,K\nonumber \\
\mu_k&\sim\text{MVN}_d(0,\omega^2I_d),\ k=1,...,K.\nonumber 
\end{align}

Direct Gibbs updates for $\{z_i\}$ and $\beta$ are hard to obtain, and we therefore leverage the Metropolis-within-Gibbs approach.

In Step 1, each $z_i$ is updated in a random order. For an arbitrary $i$ in iteration $l$, propose $\tilde{z}_{i}^l\sim\text{MVN}_d(z_{i}^l,\sigma_{z}^2I_d)$, with the acceptance probability
$$\frac{P(Y|\tilde{z}_{i}^{l},\beta)\phi_d(\tilde{z}_i^l;\{\mu_k\},\{\sigma_k^2I_d\})}{P(Y|z_{i}^{l},\beta)\phi_d(z_i^{l};\{\mu_k\},\{\sigma_k^2I_d\})}.$$

In Step 2, propose $\tilde{\beta}^l\sim \text{MVN}_d(\beta^l,\delta_{\beta}^2I_p)$, with the acceptance probability
$$\frac{P(Y|\tilde{\beta}^l,z)\phi_p(\tilde{\beta}^l;\xi,\Phi)}{P(Y|\beta^l,z)\phi_p(\beta^l;\xi,\Phi)}.$$

The full Gibbs sampling algorithm is shown in Algorithm~\ref{alg:lsm}.

\begin{algorithm}
\caption{Gibbs sampling algorithm for the latent space model}
\label{alg:lsm}
\begin{algorithmic}
\State  Denote:
\begin{align}
n_k&=\sum_{i=1}^{n}I(B(i)=k)\nonumber \\
s_k^2&=\frac{1}{d}\sum_{i=1}^{n}(z_i-\mu_k)^T(z_i-\mu_k)I(B(i)=k)\nonumber \\
\overline{z}_k&=\frac{1}{n_k}\sum_{i=1}^{n}z_iI(B(i)=k)\nonumber 
\end{align}
\For{\text{the iteration $l$}}
\State Update $\{\mu_k\}$, $\{\sigma_k^2\}$, $\{\lambda_k\}$, $\{B(i)\}$ according to:
\begin{align}
\lambda|n_{[k],k=1,...,K},\nu &\sim\text{Dirichlet}(n_1+\nu_1,...,n_K+\nu_K),\ k=1,...,K\nonumber \\
\mu_k|z,\sigma,\omega &\sim \text{MVN}_d(\frac{n_k\overline{z}_k}{n_k+\sigma_k^2/\omega^2},\frac{\sigma_k^2}{n_k+\sigma_k^2/\omega^2}I),\ k=1,...,K\nonumber \\
\sigma_k^2|\sigma_0,s_k &\sim(\sigma_0^2+ds_k^2)\text{Inv}\chi_{\alpha+n_kd}^2\nonumber 
\end{align}
$$P(B(i)=k|\lambda,z,\mu,\sigma) =\frac{\lambda_k\phi_d(z_i;\mu_k,\sigma_k^2I_d)}{\sum_{k'=1}^{K}\lambda_k\phi_d(z_i;\mu_{k'},\sigma_{k'}^2I_d)}$$ 
\State Use Metropolis-Hastings to update $\{z_i\}$ as described in Step 1
\State Use Metropolis-Hastings to sample $\beta$ as described in Step 2

\EndFor
\end{algorithmic}
\end{algorithm}

\subsection{Stopping Rule}

We use the Bayes factor as the stopping rule. Since the Bayes factor measures the strength of the model as supported by data, we compare two models -- one with community structure ($M_1$) and one without community structure ($M_0$) -- using the ratio of the model likelihoods. 
The Bayer factor increases as the evidence in the data supporting $M_1$ increases, which indicates the existence of community structure. The algorithm stops dividing the current sub-networks if the Bayes factor fails to reach a certain threshold. We refer to~\cite{jeffreys1998theory} for the selection of the cutoff value; a Bayes factor of $10$ is viewed as strong evidence in support of $M_1$ and we use $10$ as the threshold value throughout the paper. 

Direct calculation of the marginal likelihoods is difficult and therefore, we use the posterior as an approximation. Let $\Theta_0$ denote the parameter under $M_0$ and $\Theta_1$ the parameter under $M_1$. The marginal likelihood is $P(\mathbf{Y}|M_i)=\int P(\mathbf{Y}|\Theta_i, M_i) P(\Theta_i|M_i)d\Theta_i$ for an arbitrary model $M_i$, that can be approximated by Gibbs samples as $P(\mathbf{Y}|M_i)\simeq\frac{1}{L}\sum_{l=1}^{L} P(Y|\Theta_i^l,M_i)$, where $l\in[L]$ is the Gibbs iteration index. 
In practice, the question to be answered is how to specify $M_0$ and $M_1$. For simplicity, we consider a single binary division of the network. 

\subsubsection{Edge Exchangeable Model}

In the edge exchangeable model, the community structure is encoded by the $\mathcal{B}$ matrix. Assume $\mathcal{B}$ is symmetric. Let us define
$$M_0:\mathcal{B}_0=\begin{bmatrix} a& 1-a\\ 1-a & a\end{bmatrix};\ M_1:\mathcal{B}_1=\begin{bmatrix} a'& 1-a'\\ 1-a' & a'\end{bmatrix}.$$ 

Let $a\in \mathcal{A}_0$ and $a'\in\mathcal{A}_1$.  We assume that $\mathcal{A}_0\cup\mathcal{A}_1=[0.5,1]$, $\mathcal{A}_0\cap\mathcal{A}_1=\emptyset$, and $a<a'$ $\forall a\in\mathcal{A}_0$ and $a'\in\mathcal{A}_1$. One example is $\mathcal{A}_0=\{0.5\}$ and $\mathcal{A}_1=(0.5,1]$.  In this case, the goal is to evaluate whether there is no community structure in the network, i.e., whether the within community connection probability is the same as the between community connection probability. The denominator and numerator in the Bayes factor are 
$$P(\mathbf{Y}|M_0)=\int P(\mathbf{Y}|a=0.5,\Theta_{0,-a}, M_0) P(\Theta_0|M_0)d\Theta_0,$$ $$P(\mathbf{Y}|M_1)=\int \int P(\mathbf{Y}|a',\Theta_{1,-a'}, M_1) P(\Theta_1|M_1)P(a'|M_1)d\Theta_1da'.$$ 
Consider another example where $\mathcal{A}_0=[0.5, 0.6]$ and $\mathcal{A}_1=(0.6,1]$. In this case, the goal is to evaluate if the within community connection probability is greater than 0.6. The numerator stays the same as the one in the previous example, but the denominator now becomes $P(\mathbf{Y}|M_0)=\int\int P(\mathbf{Y}|a,\Theta_{0,-a}, M_0) P(\Theta_{0,-a}|M_0)P(a|M_0)d\Theta_0da$.

If the parameter space contains only a single value, the integral is calculated by sample-based approximation when fixing the within community connection probability, i.e., $a=0.5$. If the parameter space is an interval, the integral is calculated based on posterior samples such that 
$$P(\mathbf{Y}|M_i)=\int\int P(\mathbf{Y}|a_i,\Theta_{i,-a_i}, M_i) P(\Theta_{i,-a_i}|M_1)P(a_i|M_i)d\Theta_ida_i\simeq\frac{1}{L}\sum_{l=1}^{L} P(Y|\Theta_i^l,M_i),$$ 
where the distribution of $a_i$ is restricted in the corresponding parameter space $\mathcal{A}_i$ when proposing new values in the MCMC. 

\subsubsection{Stochastic Block Model}

In the stochastic block model, the community structure is reflected by the B matrix. Assume
B is symmetric. Let us define:

$$M_0:\mathcal{B}_0=\begin{bmatrix} a& b\\ b & a\end{bmatrix};\ M_1:\mathcal{B}_1=\begin{bmatrix} a'& b'\\ b' & a'\end{bmatrix}.$$ 

Let $(a,b)\in \mathcal{A}_0$ and $(a',b')\in\mathcal{A}_1$.  We assume that $\mathcal{A}_0\cup\mathcal{A}_1=[0,1]^2$, $\mathcal{A}_0\cap\mathcal{A}_1=\emptyset$, and $(a-b)<(a'-b')$, $a>b$, $a'>b'$ $\forall (a,b)\in\mathcal{A}_0$ and $(a',b')\in\mathcal{A}_1$. One example is $\mathcal{A}_0=\{(a,b)|a=b\}$ and $\mathcal{A}_1=\{(a',b')|a'>b'\}$.  In this case, the goal is to evaluate whether there is no community structure in the network, i.e., the within community connection probability is the same as the between community connection probability. 

Consider another example where $\mathcal{A}_0=\{(a,b)|a-b<0.1\}$ and $\mathcal{A}_1=\{(a,b)|a-b>=0.1\}$. In this case, the goal is to evaluate if the within community connection has 10\% higher chance to take place than between community connection.
Similar to that in the edge exchangeable model, the denominator and numerator in the Bayes factor can be approximated by the posterior samples in both examples.

\subsubsection{Latent Space Model}

Under the model specification of Section 2.1, the likelihood of the latent space model is
$$\text{logit}(Y|\{B(i)\})=\int\int P(Y,\{z_i\},\beta|\{B(i)\})dP(\beta|\xi)dP(\{z_i\}|\{\mu_k\})=\prod_{i\neq j}-\xi |\mu_{B(i)}-\mu_{B(j)}|.$$

The community structure is reflected in the mean parameter of the Gaussian mixture $\{\mu_k\}$ and the mean parameter $\xi$ of the distribution of the coefficient $\beta$ in the logistic regression. Suppose the value of $\beta$ is fixed, in which case the community structure will be determined by the $\mu_k$'s. Let $(\mu_1^*,\mu_2^*)\in \mathcal{A}_0$ and $(\mu_1',\mu_2')\in\mathcal{A}_1$.  We assume that $\mathcal{A}_0\cup\mathcal{A}_1=\mathbb{R}^2$, $\mathcal{A}_0\cap\mathcal{A}_1=\emptyset$, and $|\mu_1^*-\mu_2^*|<|\mu_1'-\mu_2'|$, $\forall (\mu_1^*,\mu_2^*)\in\mathcal{A}_0$ and $(\mu_1',\mu_2')\in\mathcal{A}_1$. 
Suppose that the goal is to evaluate whether there is community structure in the network; we may assume $\mu_1^*=\mu_2^*$ under $M_0$ and $\mu_1\neq\mu_2$ under $M_1$. Suppose the goal is to evaluate if there is strong community structure in the network; now we can restrict $|\mu_1^*-\mu_2^*|<\epsilon$ under $M_0$ and $|\mu_1'-\mu_2'|\ge\epsilon$ under $M_1$. 
Suppose $\beta$ is not fixed. In this case, we may also test if $\beta=0$ under $M_0$ in addition to comparing the $\mu^*$'s. 


\section{Simulations}
\label{s:simu}

In this section, we demonstrate our method in the three models and simulate networks for these models. We show the efficacy of the community detection framework using the Bayes factor as stopping rule. We also compare the proposed framework with other model selection criteria, including the eigenvalues in spectral clustering and information criteria.

\subsection{Simulation Set-up}

We take a node-centric approach to generate networks in the stochastic block model and the latent space model, where the number of nodes is pre-determined and the edges between node pairs are random. Denote the number of model as $n$. Suppose the number of communities $K$ is known apriori. The block assignment of an arbitrary node $i,\ i\in[n]$ in the network is sampled from a Multinomial distribution:
$$B(i)\sim\text{Multinomial}(\pi_1,...,\pi_K).$$

In the stochastic block model, the probability that a node forms an edge with another node depends on the block assignments. An edge between two nodes can be drawn from a Bernoulli distribution:
$$A_{ij}\sim\text{Bernoulli}(\mathcal{B}(B(i),B(j))),$$
where $\mathcal{B}$ is a $K$ by $K$ matrix, with each entry indicating the probability of within/between community interactions.

In the latent space model, the probability that a node forms an edge with another node depends on the latent status of nodes. Denote the latent space position of an arbitrary node $i$ as $z_i$, which can be drawn from a mixture of multivariate Gaussian distributions:
$$z_i\sim \sum_{k=1}^{K}\lambda_k \text{MVN}(\mu_k,\sigma_k^2).$$

The probability of an interaction between two nodes $i$ and $j$ is a function of the latent positions $z_i$ and $z_j$:
$$\text{logit}(P(Y_{ij}=1))=-\beta|z_i-z_j|.$$

The network data may also be generated by interaction in the edge exchangeable model. For illustration purposes, suppose $m$ interactions $E_{[m]}$ have been observed along with the block assignments. Consider a new interaction $E_{m+1}$. Suppose the number of communities $K$ and the propensity matrix $B$ are known a priori. The block assignment of the sender is sampled according to $\pi_k$ such that
$$B(S_{m+1})\sim \text{Multinomial}(\pi_1,...,\pi_K).$$

Each interaction in the network is assumed to be initiated from one block and point to another block, the probability of which is characterized by the propensity matrix $\mathcal{B}$. Given the block assignment of the sender, the block assignment of the receiver is given by
$$B(R_{m+1})|B(S_{m+1})=k\sim \text{Multinomial}(\mathcal{B}(k,1),...,\mathcal{B}(k,K)).$$

Given the block assignments of both the sender and the receiver, the next step is to select the corresponding sender and receiver nodes. We use the strategy described in the interaction-framed network~\cite{zhang2024node} where the node distribution follows a Pitman-Yor process~\cite{pitman2002combinatorial}. In this way, the sender node $S_{m+1}$ can then be drawn according to 
$$P(S_{m+1}=s|B(S_{m+1})=k)\propto
\begin{cases}
      D_{m} (s) -\alpha_{k}, & \text{if}\ s \in \mathcal{P}_{m,b} \\
      \theta_{k}+\alpha_{k} |\mathcal{P}_{m,k}|, & \text{if}\ s \not \in \mathcal{P}_{m,k}
\end{cases}
$$
where $\mathcal{P}_{m,k}$ is the set of nodes observed in the first $m$ interactions and has block assignment $b$; $|\mathcal{P}_{m,k}|$ is the cardinality of the set $\mathcal{P}_{m,k}$; $D_{m}(i), i\in \mathcal{P}_{m,k}$ is the degree of node $i$ in the first $m$ iterations; and $\alpha_{k}$ and $\theta_{k}$ are two parameters in the Pitman-Yor process. 
The intuition here is that the frequency of observing a node depends on the observed degree as well as the network properties characterized by $\alpha$ and $\theta$. The higher the observed node degree, the more likely the node is to be sampled again. We follow a similar procedure to draw the receiver node $R_{m+1}$. The entire network can be generated by repeating this procedure. We assume that nodes from different blocks are non-overlapping in the simulation.  

Intuitively, the within/between community connection strength in the stochastic block model and edge exchangeable model is determined by the the entries in $\mathcal{B}$ matrix. In contrast, in the latent space model, the within/between community connection strength is determined by the latent positions $z$ and coefficient $\beta$. In the simulation, the parameters in the corresponding generative distributions can be tuned to mimic a broad range of scenarios.

\subsection{Bayes Factor as Stopping Rule}

In the section, we show how the Bayes factor can be used as the stopping rule in our framework. We start with the simulation setting where there are two communities in the network. Our goal is to demonstrate that the algorithm is able to divide the network into two parts when community structure exists in the network. The algorithm is expected not to divide the network when there is no community structure. 

\begin{figure*}
  \includegraphics[width=\textwidth]{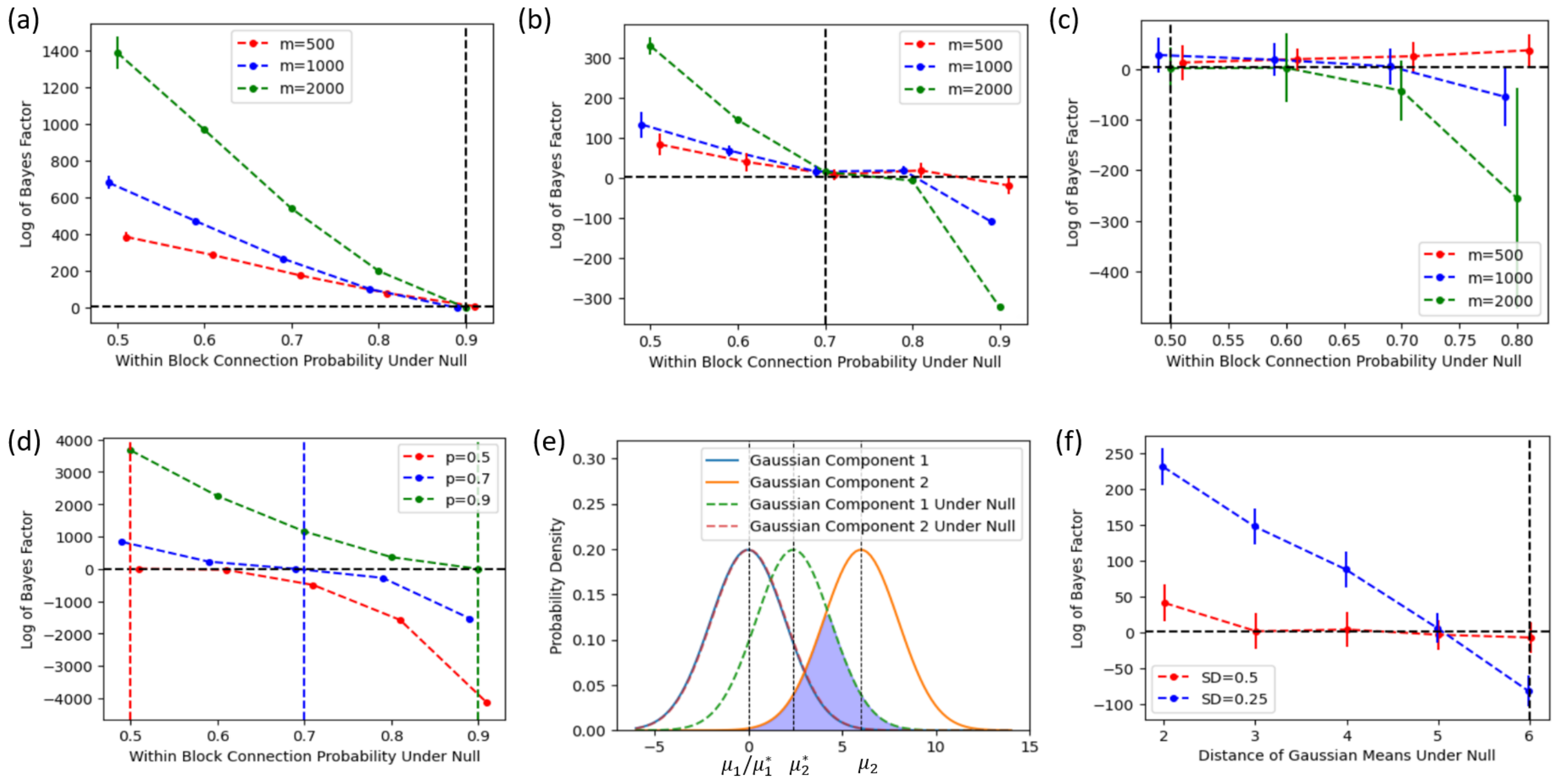}
  \caption{(a-c) Log of the Bayes factor as a function of the threshold value $t$ in the edge exchangeable model when the generative within community connection probability is (a) $a=0.9$, (b) $a=0.7$, and (c) $a=0.5$. Different colors indicate different network sizes. The horizontal dashed line indicates the cutoff value of Bayes factor. The vertical line indicates the true within community connection probability. (d) Log of Bayes factor as a function of the threshold value $t$ in the stochastic block model. (e) Illustration of the latent space model. (f) Log of Bayes factor as a function of the distance between Gaussian means in the latent space model.}
  \label{fig:1}
\end{figure*}

We start with the edge exchangeable model. Denote $\mathcal{A}_0=[0.5,t]$ and $\mathcal{A}_1=(t,1]$, where $t$ is the threshold value. We show in Figure~\ref{fig:1}(a-c) the log of the Bayes factor as a function of the threshold value $t$. We simulated three settings where the true within community connection probabilities are 0.9, 0.7, and 0.5 in the generative models. For  value of $t$ in each setting, we repeat the procedure 20 times, and construct confidence intervals for the Bayes factor. The algorithm will stop if the cutoff value falls within the interval or is greater than the Bayes factor.

In settings where community structure exists or the within community connection probability is greater than $t$, model selection criterion based on the Bayes factor is able to select the correct model at the given cutoff value. In the setting where there is no community structure or the within community connection probability is less than $t$, the algorithm will not divide the network at the given cutoff value. The accuracy of the Bayes factors increases as the number of observations increases, as expected.

In the stochastic block model, denote $\mathcal{A}_0=\{(a,b)|a-b<t\}$ and $\mathcal{A}_1=\{(a,b)|a-b\ge t\}$, where $t$ is the threshold value. Figure~\ref{fig:1}(d) shows the log of the Bayes factor as a function of the threshold value. We simulate three settings correspond to the scenarios where the true within/between community connection probability $(a,b)$ is $(0.9,0.1)$, $(0.7, 0.3)$, and $(0.5, 0.5)$. We repeat each experiment 20 times. The results show that the algorithm is able to identify the correct community structure in the network.

In the latent space model, denote $\mathcal{A}_0=\{(\mu_1^*,\mu_2^*)||\mu_1^*-\mu_2^*|<t\}$ and $\mathcal{A}_1=\{(\mu_1',\mu_2')||\mu_1'-\mu_2'|\ge t\}$. We assume $\beta$ to be fixed. Figure~\ref{fig:1}(e) shows the illustration. Denote the means of the true generative Gaussian distributions as $\mu_1$ and $\mu_2$. Align the $\mu_1^*$ under $M_0$ with $\mu_1$. Evaluating the $M_0$ is equivalent to evaluating the overlapping area between two curves -- the Gaussian distribution characterized by $\mu_2$ and $\mu_2^*$. Figure~\ref{fig:1}(f) shows the log of the Bayes factor as a function of the distance between $\mu_1^*$ and $\mu_2^*$. The true distance in the generative model is 6. We simulated two settings corresponding to the scenarios where the standard deviation is 0.25 and 0.5. The algorithm has better performance when the overlapping area is smaller. 

Out simulations show the efficacy of the Bayes factor as the stopping rule for the algorithm. We experiment on different priors as shown in Supplemental Materials. 

\subsection{Model Selection}

We show in this section how the Bayes factor can be used as a model selection criterion, and compare its performance with spectral clustering and information criteria in all three models. We use the deviance information criterion (DIC) to avoid tuning parameter issues in other information criterion. Results are shown in Table~\ref{t:one}. 

In spectral clustering, we apply the eigenvalue decomposition to the normalized Laplacian matrix, which is defined as $L=I-D^{-1/2}AD^{-1/2}$, where $D$ is the diagonal degree matrix. The number of communities is determined by the change point in the scale of the eigenvalues. We compute the DIC as $DIC=D(\overline{\theta})+2p_D$,
where $D(\theta)=-2\log L(y|\theta)$ and $p_D=E_{\theta|y}[D]-D(E_{\theta|y}(\theta))$, where $\theta$ is the model parameter.
Note that the spectral clustering approach is deterministic, while the Bayes factor and the information criterion are stochastic. Results from the Bayes factor and DIC in Table~\ref{t:one} are the mostly frequent number of communities after repeating the experiments 10 times. The actual times of correctly identifying the true value in the 10 repeats are shown in parentheses.

\begin{table}
\caption{Model Selection Results Using Different Selection Criteria}
\label{t:one}
\begin{center}
\begin{tabular}{lcccccc}
\hline
Model & Setting & \multicolumn{1}{c}{$\text{Size}$} & \multicolumn{1}{c}{$\text{Truth}$} &\multicolumn{1}{c}{$\text{Bayes Factor}$} &  \multicolumn{1}{c}{$\text{Spectral Clustering}$} & 
\multicolumn{1}{c}{$\text{DIC}$} \\ \hline
EE& Simu 1 & 1,000 & 4 & 4 (10/10) & $>$ 20 & 2 (4/10)\\
& Simu 2 & 1,000 & 4 & 4 (9/10) & $>$ 20 & 2 (0/10)\\
& Simu 3 & 2,000 & 4 & 4 (9/10)& $>$ 20 & 4 (5/10)\\
& Simu 4 & 2,000 & 8 & 8 (8/10)& $>$ 20 & 4 (0/10)\\
\hline
SBM & Simu 1 & 100 & 2 & 2 (10/10) & -- & -- \\
& Simu 2 & 200 & 4 & 4 (10/10) & -- & -- \\
\hline
LSM & Simu 1 & 100 & 2 & 2 (10/10) & -- & 2 (10/10)\\
& Simu 2 & 200 & 4 & 4 (8/10) & -- & 4 (5/10)\\
\hline
\end{tabular}
\end{center}
\end{table}

We simulated four different settings in the edge exchangeable model. In Setting 1 and Setting 2, the network contained 1,000 interactions and the within community connection probability was set to 0.9 and 0.7, respectively. 
In Settings 3 and 4, the network contained 2,000 interactions and the within community connection probability was 0.9, but the true number of underlying communities was set to 4 and 8, respectively. Our method outperforms the comparands in all settings.

We further simulated two settings for the stochastic block model. In Settings 1 and 2, the network contained 100 and 200 nodes, respectively. The within community connection probability was 0.9. We did not find clear cut-off eigenvalues when applying the spectral clustering approach to the data. The DIC kept increasing with $K$, indicating a preference for complex models. Lastly, we simulated two different settings using the latent space model. In Settings 1 and 2, the network contained 100 and 200 nodes, respectively. The Gaussian means were $\mu_1=0.0,\text{and\ } \mu_2=6.0$; and $\mu_1=[0.0, 0.0],\ \mu_2=[0.0, 6.0],\ \mu_3=[6.0, 0.0],\text{and\ } \mu_4=[6.0, 6.0]$, respectively. We assumed a constant standard deviation of 0.25. We did not find clear cut-off eigenvalues using spectral clustering in this case either. Our method performs slightly better than the DIC.

In conclusion, the Bayes factor method outperforms the other methods in most settings, especially when only few observations are available and within-community connection is weak. 


\section{Application to Stroke Patient Transfer Network}

\subsection{Data Overview}

In the section, we apply our method to stroke patient transfer network data. The data was collected from all 342 non-federal California hospitals, including around 350,000 stroke patients from 2010 to 2020. Access to timely interventions is critical in stroke care, and there may be only a few hours to prevent disability or preserve life. Therefore, patients are sometimes transferred to hospitals to access certain treatments and procedures that are typically available only in select hospitals, such as those affiliated with universities. 

We constructed the network by treating the hospitals as nodes. An interaction was drawn between two hospitals if there was at least one patient transferred from one to the other. We only considered transfers that were initiated from emergency departments and happened within 24 hours of initial admission. From 2010 to 2020, a total of 17,562 encounters/transfers was reported in the data. The final dataset consisted of 335 hospitals, with an average of 4.9 patients transferred out per sending hospital and 6.5 patients accepted per receiving hospital. We stratified the data by year, which led to 11 non-overlapping networks.

\begin{figure*}
  \includegraphics[width=\textwidth]{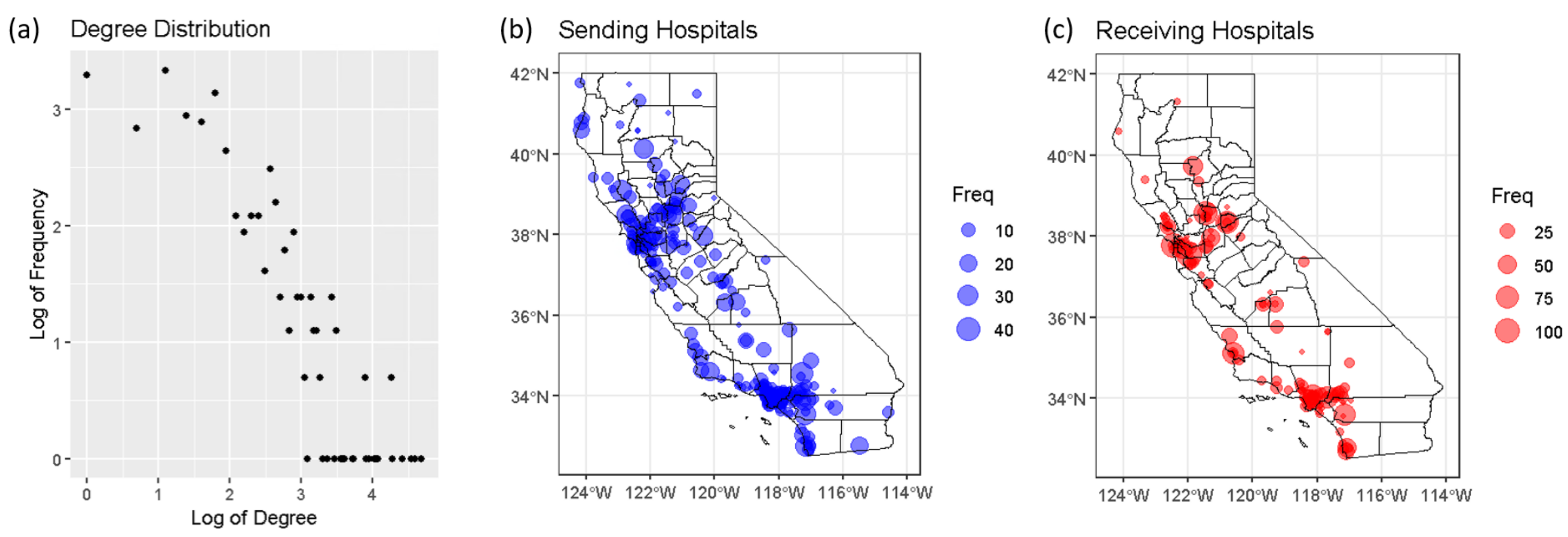}
  \caption{(a)Degree distribution of both sending and receiving hospitals in Year 2020; (b) Geographical distribution of sending hospitals in Year 2020. Size of the dots indicate the number of transfers; (c) Geographical distribution of receiving hospitals in Year 2020. Size of the dots indicate the number of transfers}
  \label{fig:3}
\end{figure*}

Figure~\ref{fig:3} shows the degree distribution of the transfer network in 2020, which is highly skewed and has few high-degree hospitals and many low-degree hospitals, which fits the underlying assumptions of the edge exchangeable model. Figures~\ref{fig:3}b and~\ref{fig:3}c show the geographical distribution of the sending and receiving hospitals. Overall, the sending hospitals are scattered across all regions, while the receiving hospitals tend to be concentrated at regional centers.

\subsection{Fit to the Data}

We assumed $M_0$: $a=0.5$ and fit the model. Given the time sensitive nature of stroke interventions, the geographical distance and hospital affiliations may have clinical implications in the treatment of stroke patients. We incorporated these application specific covariates on extra information when fitting the data (see Supplemental Materials for details). 

A closer look at the actual transfers within and between communities is shown in Figure~\ref{fig:4}. The observed block diagonal structure means that the within community transfers are more likely than the between community transfers, which indicates that the proposed method is able to identify the latent structure of the network. We examined the geographical distributions of the detected communities.  Hospitals that are geographically close tend to belong to the same community. 





\subsection{Evaluation of Results}

\begin{figure*}
  \includegraphics[width=\textwidth]{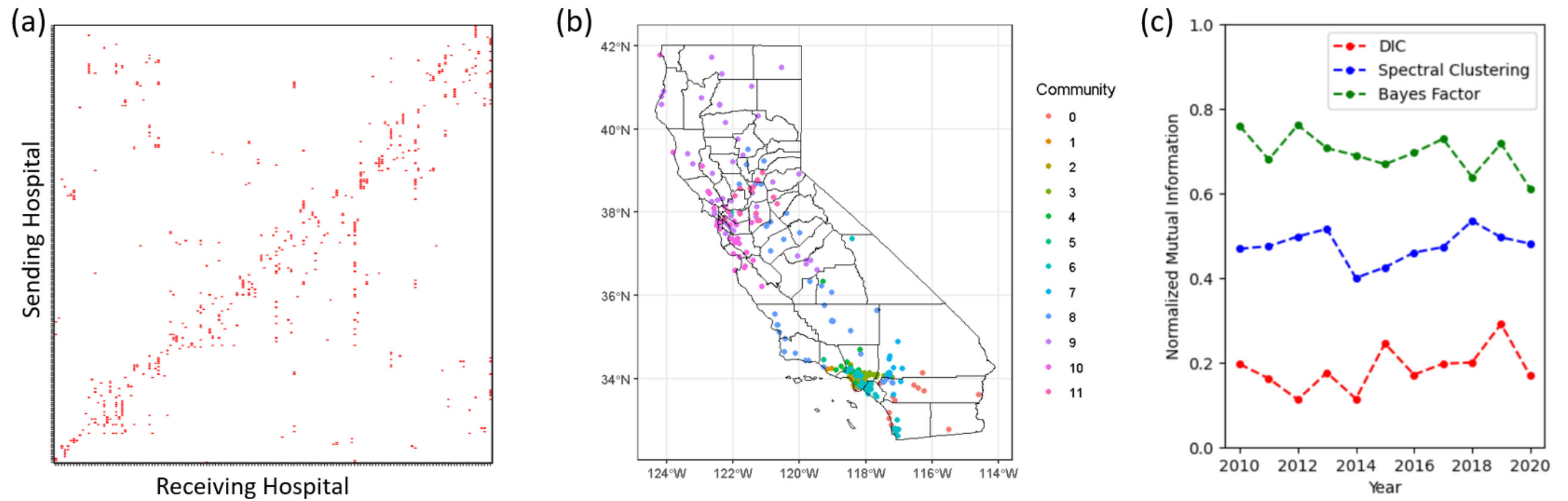}
  \caption{(a) Transfers within/between hospitals. Red indicates at least one transfer from a sending hospital (y-axis) to a receiving hospitals (x-axis). Both sending and receiving hospitals are ordered by community labels. (b) Geographical distribution of hospitals (dots) and the detected communities indicated by color. (c) Normalized mutual information (NMI) in different years.}
  \label{fig:4}
\end{figure*}

In this section, we evaluate the quality of the detected communities. 
We randomly split the data into two parts by interaction and fit the model separately to the two parts and checked the consistency of the two sets of labels.
To quantify the level of consistency, we used the normalized mutual information (NMI) defined as
$$\text{NMI}(W,C)=\frac{2I(W,C)}{H(W)+H(C)},$$
where $C$ and $W$ are the two sets of labels in the two parts of the data; $H$ is the entropy, and $I$ is the mutual information. For each comparison, NMI ranges between 0 and 1: $\text{NMI}=1$ means the two sets of labels are identical, and 0 means there is no shared information between two sets of labels. We compare our proposed framework with other methods. Results are shown in Figure~\ref{fig:4}. 

Note that given the sparsity and the small size of the network, it is in general difficult to obtain consistent results, but our framework nevertheless gives reasonable results and consistently perform better than the spectral clustering and the deviance information criterion.

\section{Discussion}
\label{s:discuss}

In this paper, we proposed a new method for community detection through recursive bi-partitioning of the network. Our framework is based upon the Bayes factor and is applicable to a wide range of models. We demonstrated our method using the stochastic block model, edge exchangeable model, and latent space model using simulated data. Compared to the other methods, our method had good performance in identifying the correct communities in the network. We also applied the method to the stroke patient transfer network, where we identified the underlying hospital clusters that could be further used for investigating regional disparities in health care access. 

Our method has several limitations. Though not common, in some cases the algorithm can pick the wrong model, especially when the sample size is small. Its performance in the stochastic block model is overall stable. However, the performance in the edge exchangeable is affected by the number of communities. We hypothesize this is due to the presence of more low-degree nodes in a larger number of communities. To overcome this problem, we can either increase the sample size or utilize external information as the prior. In addition, in the latent space model, we need to put several restrictions on the model parameters, e.g., fix $\beta$. There is room for future method development.

\printbibliography




\end{document}